\documentclass[10pt, onecolumn]{article}

\usepackage{cite}
\usepackage[cmex10]{amsmath}
\usepackage{fixltx2e}
\usepackage{datetime}
\usepackage{graphicx}
\usepackage{stfloats}
\usepackage{color}
\usepackage[T1]{fontenc}
\usepackage[utf8]{inputenc}
\usepackage{authblk}
\usepackage{amssymb}
\usepackage{algorithm}
\usepackage{pstricks,epsfig}
\usepackage{rotating}
\usepackage{subfig}
\usepackage{tikz}
\usepackage{enumerate}
\usepackage{tabularx}
\usepackage{verbatim}
\usepackage{pgf}
\usepackage{url}
\usepackage{geometry}
\usepackage{times}
\newcounter{lastnote}

\title{Laser light-field fusion for wide-field lensfree on-chip phase contrast nanoscopy}

\author{Farnoud~Kazemzadeh$^{1,\ast}$ and Alexander~Wong$^1$\\
\normalsize{$^{1}$Department of Systems Design Engineering, University of Waterloo}\\ \normalsize{200 University Avenue West, Waterloo, Ontario, Canada, N2L 3G1}\\
\vspace{0.5cm}
\normalsize{$^\ast$ To whom correspondence should be addressed; E-mail: fkazemzadeh@uwaterloo.ca}}

\date{}

\begin{document}

\baselineskip24pt

\maketitle

\begin{abstract}
Wide-field lensfree on-chip microscopy, which leverages holography principles to capture interferometric light-field encodings without lenses, is an emerging imaging modality with widespread interest given the large field-of-view compared to lens-based techniques. In this study, we introduce the idea of laser light-field fusion for lensfree on-chip phase contrast nanoscopy, where interferometric laser light-field encodings acquired using an on-chip setup with laser pulsations at different wavelengths are fused to produce marker-free phase contrast images of superior quality with resolving power more than five times below the pixel pitch of the sensor array and more than 40\% beyond the diffraction limit. As a proof of concept, we demonstrate, for the first time, a wide-field lensfree on-chip instrument successfully detecting 300 nm particles, resulting in a numerical aperture of 1.1, across a large field-of-view of $\sim$ 30 mm$^2$ without any specialized or intricate sample preparation, or the use of synthetic aperture- or shift-based techniques.
\end{abstract}
%
\section*{Introduction}
Phase contrast microscopy was introduced as a means of observing objects with light transparent properties~\cite{Zernike42}. Bright-field microscopy of such objects tend to produce low contrast images since the probing light and the sample being imaged do not undergo a very strong interaction. Phase contrast microscopy, however, allows for visualization of the optical path length differences that the probing light would experience as the result of the interaction with the sample, thereby producing high contrast images of such samples. Despite the aforementioned benefit of phase contrast microscopy, it nevertheless suffers from the same shortcomings of bright-field microscopy such as limited field-of-view (FOV), limited resolution capability, and high complexity of the design and operation of the instrument.

Wide-field lensfree on-chip microscopy, where holography principles are used to capture interferometric light-field encodings without lenses, has become an interesting pathway toward addressing the shortcomings of lens-based techniques~\cite{Frentz10,Su13,Mudanyali10,Sheng06,Kiss13,Mico06,Mico10,Granero10,Leon08,Luo15, Bishra11,Greenbaum13,Isikman12,Isikman10,Greenbaum13a,Noom1,Noom2,Wong15,Kazemzadeh1, Kazemzadeh2}. At its core, these lensfree on-chip instruments are conceptually simple and offer extremely large FOV in comparison to lens-based instruments. The resolution of such instruments is inherently limited by the pixel pitch of the detectors used, which is currently limited by the fabrication technologies to $\sim$ 1 $\mu$m. The quest for higher resolving power has led to numerous synthetic aperture- or lateral shift-based techniques~\cite{Mico06,Luo15,Mico10,Pelagotti12,Ralston07,Granero10} for increasing resolving power beyond pixel pitch, though such techniques tend to increase the hardware complexity and decrease performance tolerance of the instrument. More recently, multi-wavelength illumination-based techniques~\cite{Noom1,Kazemzadeh1,Kazemzadeh2,Luo16} showed considerable promise for increasing the resolving power beyond pixel pitch without the level of complexity of synthetic aperture- or lateral shift-based techniques.

A particular area of interest for employing wide-field lensfree on-chip microscopy is for nano-scale imaging.  Super-resolution microscopy~\cite{Hell94,Betzig92}, atomic force microscopy~\cite{Binnig86}, and electron microscopy~\cite{Azubel14} have been the mainstays for nano-scale imaging but are limited by very high instrumentation cost and complexity, narrow FOV, and low imaging throughput due to the complexity of the imaging process.  The ability to harness the principles behind wide-field lensfree on-chip microscopy for nano-scale imaging, which we will refer to as wide-field lensfree on-chip nanoscopy, holds considerable promise for greatly reducing imaging cost and complexity while improving image throughput and FOV, thus enabling comprehensive, large scale studies to be more readily achieved.  Recent results are quite promising~\cite{Mudanyali13}, demonstrating the ability to distinguish nanoparticles of various sizes.  However, such techniques require specialized and intricate sample preparation (e.g., \cite{Mudanyali13} requires the use of biocompatible wetting films to self-assemble aspheric liquid nanolenses around individual nanoparticles) which significantly increases imaging complexity.  As such, a method to achieve wide-field lensfree on-chip nanoscopy to facilitate for the detection of objects at the nanometer-scale without the need for additional specialized sample preparation, or the use of synthetic aperture- or lateral shift-based techniques is highly desired.

\begin{figure*}[!t]
	\centering
    \includegraphics[width=1\linewidth]{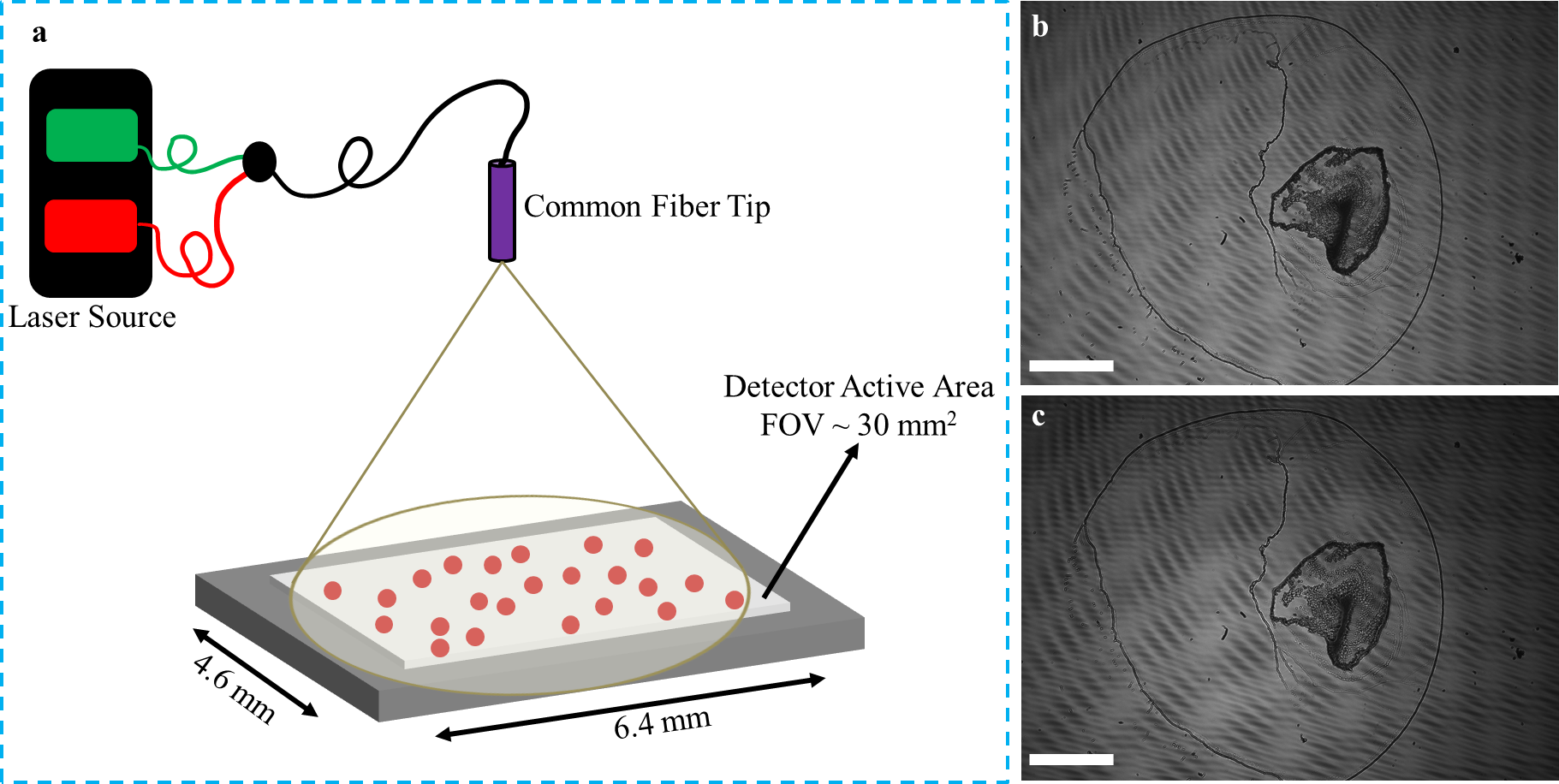}
	\caption{\textbf{The proposed instrument along with the imaging results.} (\textbf{a}) The schematic of the instrument showing the single-mode fiber optic cable which is used by two-channel pulsed laser light source; the size of the detector array which is the total field-of-view of the instrument is noted. (\textbf{b} and \textbf{c}) The captured interferometric light-field encoding ($\lambda_{1}=531.9$ nm (\textbf{b}) and $\lambda_{2}=638.3$ nm (\textbf{c})), with the scale bars denoting 1 mm.}
	\label{fig1}
\end{figure*}

In this study, we introduce the idea of laser light-field fusion phase contrast nanoscopy. Inspired by our earlier, preliminary exploration into spectral light-field fusion~\cite{Kazemzadeh1,Kazemzadeh2}, laser light-field fusion phase contrast nanoscopy involves the acquisition and fusion of interferometric laser light-field encodings using a lensfree, on-chip setup with laser pulsations at different wavelengths to produce marker-free phase contrast images, as shown in Fig.~\ref{fig1}.  The proposed instrument allows us to achieve a resolving power below the pixel pitch of the sensor array as well as the wavelength of the probing light source, beyond the diffraction limit.  As proof of concept, we demonstrate, for the first time, a lensfree on-chip instrument capable of detecting 300 nm particles across a large field-of-view of $\sim$ 30 mm$^2$ without utilization of specialized or intricate sample preparation, or the use of synthetic aperture- or lateral shift-based techniques.

\section*{Results}
\begin{figure*}[!h]
	\centering
    \includegraphics[width=1\linewidth]{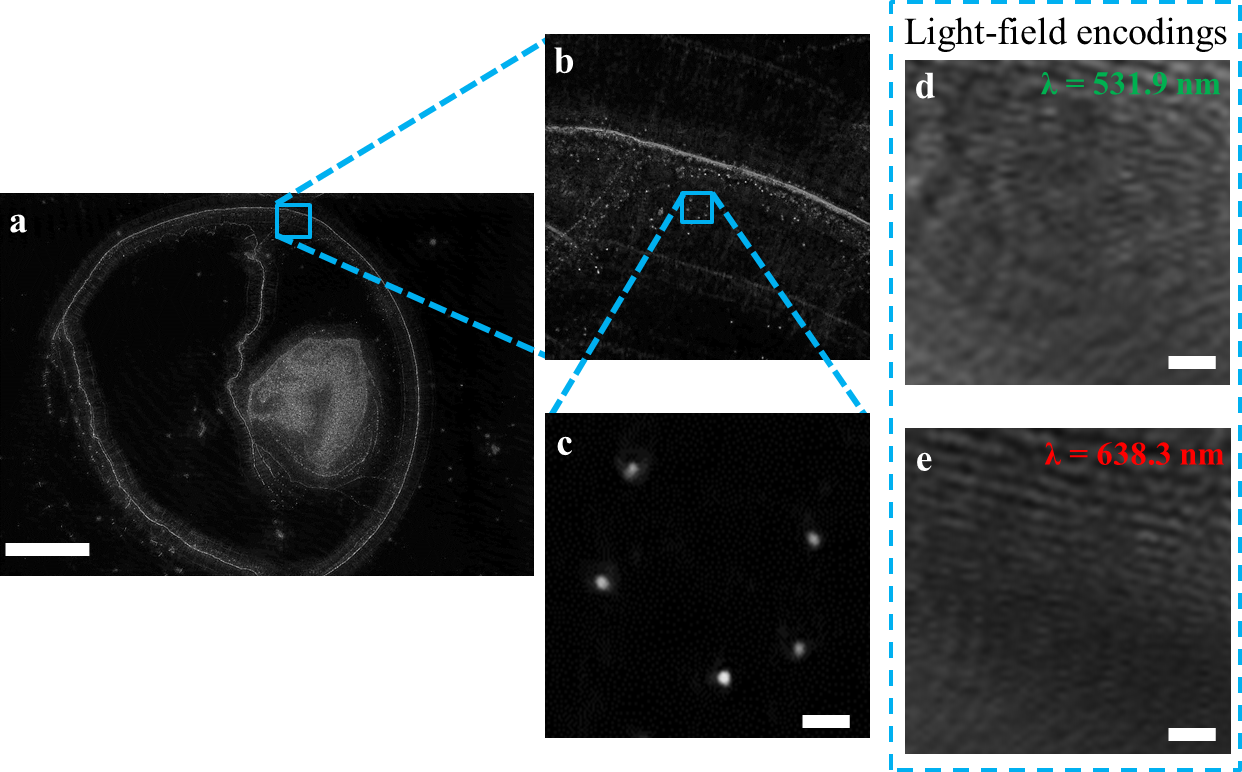}
	\caption{\textbf{The reconstructed phase contrast nanoscopy image of 500 nm nanospheres.} (\textbf{a})The full FOV image of the detector, scale bar denotes 1 mm. (\textbf{b}) A zoomed-in region of \textbf{a}. (\textbf{c}) Further zoomed-in of a specific region in \textbf{b} showing a 'U' arrangement of 500 nm nanospheres.  (\textbf{d} and \textbf{e}) The captured light-field encodings at $\lambda_{1}=531.9$ nm (\textbf{d}) and $\lambda_{2}=638.3$ nm (\textbf{e}) used to obtain \textbf{c}. The scale bars in \textbf{c, d, e} all denote 2 $\mu$m.}
	\label{fig2}
\end{figure*}

The capability of the proposed instrument in resolving objects at the nanometer scale is first demonstrated in Fig.~\ref{fig2}, where a phase contrast image of a sample consisting of polystyrene nanospheres (Fluoresbrite, Polysciences, Inc., USA) is shown in Fig.~\ref{fig2}a.  Upon closer examination, through two zoom levels, Fig.~\ref{fig2}b and c, a phase contrast image of an isolated cluster of five 500 nm nanospheres are discovered, arranged in a `U' formation. The light-field encoding captured at $\lambda_{1}=531.9$ nm and $\lambda_{2}=638.3$ nm are shown in Fig.~\ref{fig2}d and e, respectively.

\begin{figure*}[!t]
	\centering
    \includegraphics[width=1\linewidth]{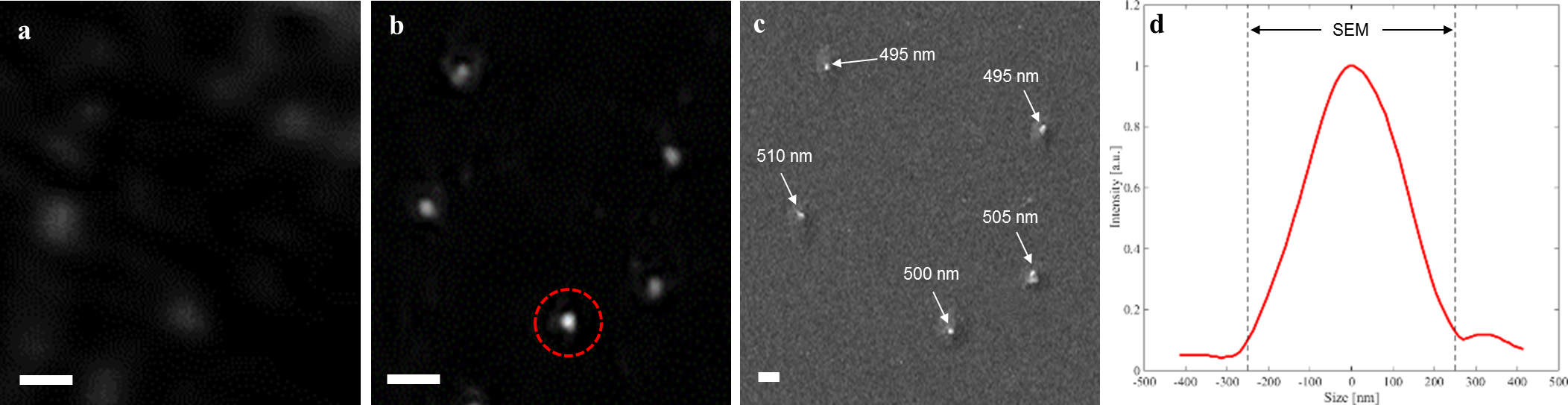}
	\caption{\textbf{Imaging of a collection of 500nm nanospheres with SEM validation.} (\textbf{a}) The phase contrast image containing five nanospheres arranged in a 'U' formation obtained using the reference lensfree on-chip instrument capturing interferometric light-field encodings at $\lambda=531.9$ nm.  (\textbf{b}) The phase contrast image obtained using the proposed instrument containing five nanospheres arranged in a 'U' formation. (\textbf{c}) The SEM image of the corresponding region with the size of the nanospheres noted on the image (with the smallest nanosphere measured at 495 nm in size). (\textbf{d}) The cross-section profile of the nanosphere, highlighted with a red dashed circle in \textbf{b}, using the our instrument accompanied by the confirmed SEM nanosphere size. The scale bars in \textbf{a},\textbf{b}, and \textbf{c} denote 2 $\mu$m.}
	\label{fig3}
\end{figure*}

Scanning electron microscopy (SEM) (MERLIN, Carl Zeiss AG, Germany) was used to verify the imaging results captured by our instrument, as shown in Fig.~\ref{fig3}. The corresponding location of the `U' formation of the nanospheres are shown using a reference lensfree on-chip instrument capturing interferometric light-field encodings at $\lambda=531.9$ nm (see Fig.~\ref{fig3}a), the proposed laser light-field fusion phase contrast nanoscopy instrument (see Fig.~\ref{fig3}b), and SEM (see Fig.~\ref{fig3}c).  The sizes of the nanospheres are noted on the SEM image which are within the manufacturer's fabrication tolerance of 500$\pm$3\% nm~\cite{Polysciences}.  A number of observations can be made from Fig.~\ref{fig3} with regards to the reference lensfree on-chip instrument and the proposed laser light-field fusion phase contrast nanoscopy instrument.  It can be observed that the same nanospheres observed by the SEM cannot be detected using the reference lensfree on-chip instrument capturing interferometric light-field encodings at $\lambda=531.9$ nm.  On the other hand, it can be observed that the same nanospheres observed by the SEM have been detected with the proposed laser light-field fusion phase contrast nanoscopy instrument, thus demonstrating its ability to detect particles that are 495 nm in size as well as achieve superior performance to the reference lensfree on-chip instrument.  Furthermore, it can be observed that differentiating the size variability of these nanospheres is beyond the capability of the proposed instrument as such variability is well beyond its resolving power.

It can also be observed that the image quality achieved using the proposed laser light-field fusion phase contrast nanoscopy instrument is high at a detection signal-to-noise ratio (SNR) of 33.35 dB, achieved by only capturing a single interferometric light-field encoding at a given wavelength in the current setup. Fig.~\ref{fig3}d shows the cross-sectional profile of the nanoparticle highlighted by the red circle in Fig.~\ref{fig3}b as well as the SEM confirmed profile.   It can be observed that the discrepancy between the measurements made by the SEM and the proposed instrument is negligible, thus demonstrating the fidelity and performance capabilities of the proposed instrument.

\begin{figure*}[!t]
	\centering
    \includegraphics[width=1\linewidth]{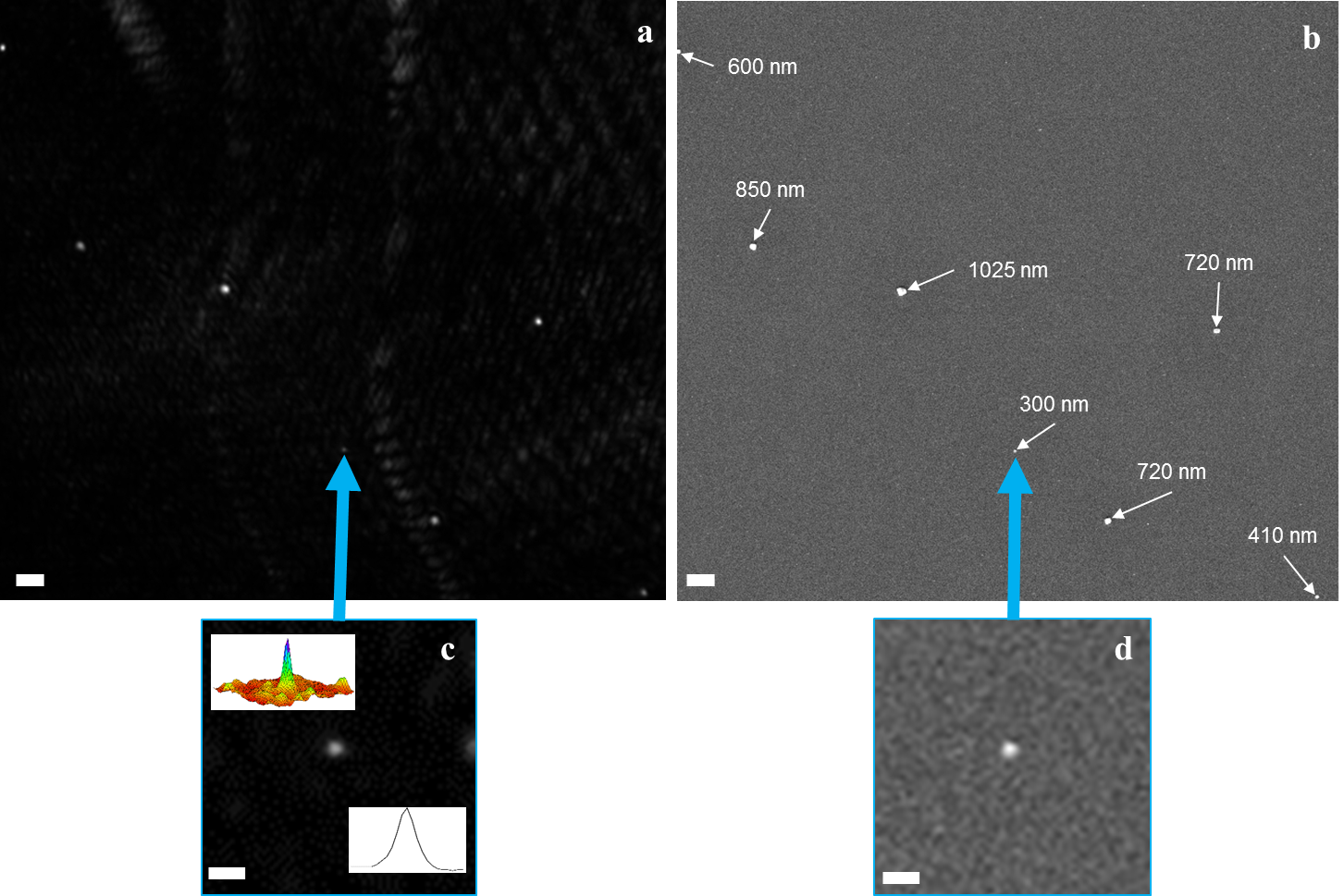}
	\caption{\textbf{Imaging of a collection of particles of different sizes with SEM validation.} (\textbf{a}) A select location of a $\sim$ 30 mm$^2$ FOV acquired by the proposed instrument containing seven particles of varying sizes. (\textbf{b}) The SEM image of the same FOV. (\textbf{c}) The zoomed-in image of the 300 nm particle with the cross-sectional profile and the surface intensity map inset. (\textbf{d}) The SEM verification image of the 300 nm particle. The scale bar in \textbf{a} and \textbf{b} denote 3 $\mu$m and in \textbf{c} and \textbf{d} denote 1 $\mu$m.}
	\label{fig4}
\end{figure*}

The capability of the proposed instrument was further demonstrated using a sample consisting of a mixture of particles of different sizes that was prepared for imaging. A zoomed-in region of this sample is shown in Fig.~\ref{fig4}a and b imaged by the proposed instrument and an SEM, respectively. There are six different sizes of particles in this region varying from 300 nm to $\sim$ 1 $\mu$m. The size of these particles are verified using the SEM and the corresponding particles are observed using the proposed instrument. A number of observations can be made from Fig.~\ref{fig4}.  It can be observed that the smallest particle, which is 300 nm in size, is detected using both the SEM and the proposed laser light-field fusion phase contrast nanoscopy instrument, with Fig.~\ref{fig4}c and d showing this particle in isolation for the proposed instrument and SEM, respectively. It can also be observed from the cross-sectional profile and the intensity surface of the particle (see inset of Fig~\ref{fig4}c) that a positive detection of the 300 nm particle is demonstrated, with a detection SNR of 29.61 dB. The imaging NA as defined by this detection is 1.1, and the improved imaging resolving power of the proposed instrument is more than five times smaller than the pixel pitch of the detector used. Furthermore, it can be observed that the size variation of different sized particles are explicitly reflected in the phase contrast image, as their imaged size tends to reflect the actual size.  These results demonstrate as a proof-of-concept the capabilities of the proposed instrument, which is beyond existing lensfree on-chip instruments that do not require any specialized or intricate sample preparation, or the use of synthetic aperture- or shift-based techniques.

%
\section*{Discussion}
We introduced an wide-field lensfree on-chip phase contrast nanoscopy instrument with resolving power beyond the diffraction limit, below the wavelength of the probing illumination source. As proof-of-concept, the capability of the proposed instrument to surpass the pixel pitch resolution limit of the detector is demonstrated by imaging particles which are 300 nm in size, which is more than five times smaller than the pixel pitch of the detector used (1.67 $\mu$m).  These experimental results demonstrate that the unique diffraction behaviour captured in the acquired interferometric laser light-field encodings at different wavelengths can be leveraged through a fusion process to achieve improved image quality and resolving power in a lensfree on-chip instrument beyond that can be achieved using a single wavelength.

It is important to note that the proposed wide-field lensfree on-chip instrument does not require specialized sample preparation, or the use of synthetic aperture- or lateral shift-based techniques to accomplish nanometer-scale resolving power, with the current setup achieving an NA of 1.1.  What this implies is that the proposed instrument has low instrumentation complexity and cost, and is easy to operate and maintain, thus allowing for democratization and proliferation of such instruments in healthcare, industry, education, and research.  The interferometric light-field encoding acquisitions made by the proposed instrument take less than $\sim$ 2ms (equivalent to > 500 frames per second), thus enabling observations of highly time-resolved dynamic systems or transient phenomena at the nanometer scale, such as for the study of motion dynamics of colloidal nanoparticles.

A limitation with the setup of the proposed instrument used in this current study is that the number of wavelengths used is restricted to two different laser wavelengths, which can be addressed in future studies through the incorporation of a tunable laser to potentially enable further improvements in image quality and resolving power.

\section*{Materials and Methods}
\textbf{Imaging apparatus}. The proposed wide-field lensfree on-chip phase contrast nanoscopy instrument used for this study (shown in Fig.~\ref{fig1}) can be described as follows.  A two-channel pulsed laser light source is used, with central wavelengths at $\lambda_{1}=531.9$ nm and $\lambda_{3}=638.3$ nm, and the spectral bandwidth being $\sim \pm$0.5 nm.  The pulsed laser light source was programmed to pulse with an alternating wavelength pulse sequence, where the duration of the pulsations are configured such that the signal observed on the detector is maximized while mitigating pixel saturation.  The pulsed laser light source is coupled to a single-mode fiber optic cable to illuminate the sample. The detector readout is synchronized to the pulse sequence for rapid and seamless interferometric light-field encoding acquisitions at the two wavelengths of the laser light source. The exposure time was < 1 ms for each wavelength, resulting in interferometric light-field encoding acquisitions using the proposed instrument taking less than $\sim$ 2 ms (equivalent to > 500 frames per second)

The sample being imaged is placed on a \#1 microscope cover slip with a thickness of $\sim$ 145 $\mu$m, which sits directly on the detector. Interferometric light-field encoding acquisitions of the sample being imaged at different wavelengths are made by the detector using a 3840$\times$2748 pixel CMOS sensor array with a pixel pitch of 1.67 $\mu$m.  The FOV of the device is determined based the active sensor size and is $\sim$ 30 mm$^2$.  The captured interferometric light-field encoding, denoted by $g_{x,y,\lambda}$ with $\lambda$ denoting wavelength, is then sent to the digital signal processing unit, where numerical laser light-field fusion is performed to reconstruct a fused phase contrast image $r_{x,y,z}$.

In this study, the measurement instrument was characterized based on a number of acquisitions of point source targets to determine the aberration transfer function at each wavelength (denoted by $H^{a}_{\lambda}$) of the pulsed laser light source to account for differences at different wavelengths.

For comparison purposes, a reference lensfree on-chip instrument capturing interferometric light-field encodings at $\lambda=531.9$ nm using the aforementioned imaging apparatus (with a single-channel laser light source is used instead) was also evaluated in this study.

\textbf{Image Reconstruction}. The numerical laser light-field fusion performed on the digital signal processing unit to reconstruct fused phase contrast images from measurements made by the proposed instrument can be described as follows.  The captured interferometric laser light-field encoding $g_{x,y,\lambda}$ encapsulates unique diffraction behaviour at different wavelengths $\lambda$, which we will leverage to achieved improved image quality beyond that can be achieved using a single wavelength.  Let us formulate a fused laser object light-field $q_{x,y,z}$ as a subspace projection of the laser object light-field $f_{x,y,z,\lambda}$,
\begin{equation}
	q_{x,y,z} = \int v_{\lambda}f_{x,y,z,\lambda} d\lambda,
\label{super}
\end{equation}
\noindent where $v_{\lambda}$ denotes the $\lambda$-associated coefficient of the largest eigenvector of the correlation matrix of $f_{x,y,z,\lambda}$, thus taking into account the correlation structure across $\lambda$.  Given that the instrument described captures $g_{x,y,\lambda}$, not $f_{x,y,z,\lambda}$, one must devise a mechanism to numerically estimate $q_{x,y,z}$ given $g_{x,y,\lambda}$.

Let us now model the laser object light-field $f_{x,y,z,\lambda}$ and the interferometric laser light-field encoding $g_{x,y,\lambda}$ as probability distributions.   To estimate $q_{x,y,z}$, we now wish to compute a subspace projection of the most probable laser object light-field $f_{x,y,z,\lambda}$ given $g_{x,y,\lambda}$, with \emph{a priori} knowledge related to $f_{x,y,z,\lambda}$ and the aberration transfer function ($H^{a}_{\lambda}$), as well as the Rayleigh-Sommerfeld diffraction transfer function ($H^{d}_{z,\lambda}$),
\begin{equation}
	\hat{q}_{x,y,z} = \int v_{\lambda}\left\{{\rm argmax}_{f_{x,y,z,\lambda}}~p\left(g_{x,y,\lambda} | f_{x,y,z,\lambda}\right)p(f_{x,y,z,\lambda})\right\}d{\lambda},
\label{MAP}
\end{equation}
\noindent where $p\left(g_{x,y,\lambda} | f_{x,y,z,\lambda}\right)$ denotes the likelihood of $g_{x,y,\lambda}$ given $f_{x,y,z,\lambda}$ and $p(f_{x,y,z,\lambda})$ denotes the prior of $f_{x,y,z,\lambda}$.  According to quantum photon emission statistics, one can express $p\left(g_{x,y,\lambda} | f_{x,y,z,\lambda}\right)$ as
{
\begin{equation}
	p\left(g_{x,y,\lambda} | f_{x,y,z,\lambda}\right) = \prod_{x \in X}\prod_{y \in Y}\prod_{z \in Z} \frac{\left(\mathfrak{F^{-1}}\left\{\frac{H^{a}_{\lambda}}{H^{d}_{z,\lambda}}\mathfrak{F}\left\{f_{x,y,z,\lambda}\right\}\right\}\right)^{{g_{x,y,\lambda}}}e^{-\left(\mathfrak{F^{-1}}\left\{\frac{H^{a}_{\lambda}}{H^{d}_{z,\lambda}}\mathfrak{F}\left\{f_{x,y,z,\lambda}\right\}\right\}\right)}}{{g_{x,y,\lambda}}!}
\label{likelihood}
\end{equation}
}
\noindent where $\mathfrak{F}$ and $\mathfrak{F^{-1}}$ denotes the forward and inverse Fourier transform, respectively.  Modeling $f_{x,y,z,\lambda}$ as a nonstationary process, one can express $p\left(f_{x,y,z,\lambda}\right)$ as
\begin{equation}
	p\left(f_{x,y,z,\lambda}\right) = \prod_{x \in X}\prod_{y \in Y}\prod_{z \in Z} e^{-\frac{\left(f_{x,y,z,\lambda}-{E}(f_{x,y,z,\lambda})\right)^2}{2 \tau^2}}.
\label{prior}
\end{equation}

\noindent where ${E}(f_{x,y,z,\lambda})$ denotes the nonstationary expectation and $\tau^2$ denotes the variance.  To reconstruct the fused phase contrast image $r_{x,y,z}$, the computed $\hat{q}_{x,y,z}$ is phase-shifted by 90$^{o}$ at the zeroth frequency, and the amplitude of this phase-shifted $\hat{q}_{x,y,z}$ is then taken as $r_{x,y,z}$~\cite{Zernike42}.

An expectation maximization for MAP estimation is used to solve Eq.~\ref{MAP}, and is performed until convergence.



\section*{Acknowledgments}
The authors would like to thank Dr. Chao Jin and Prof. Monica Emelko from department of Civil and Environmental Engineering at University of Waterloo for providing the samples used in this study.

\section*{Funding}
This work was supported by the Natural Sciences and Engineering Research Council of Canada, Canada Research Chairs Program, and the Ontario Ministry of Research and Innovation.

\section*{Author contributions}
F.K. and A.W. contributed equally to this study.

\section*{Competing Interests}
The authors declare that they have no competing interests.

\section*{Data and Material Availability}
All data related to this paper may be requested from the authors by emailing either fkazemzadeh@uwaterloo.ca or alexander.wong@uwaterloo.ca 

\pagebreak

\clearpage

\clearpage

\clearpage

\begin{thebibliography}{99}
\bibliographystyle{Science}

\bibitem{Zernike42} F. Zernike, Phase contrast, a new method for the microscopic observation of transparent objects, Physica \textbf{9}, 686-698 (1942).

\bibitem{Frentz10}  Z. Frentz, S. Kuehn, D. Hekstra, S. Leibler, Microbial population dynamics by digital in-line holographic microscopy, Rev. Sci. Instrum. \textbf{81} 084301 (2010).

\bibitem{Su13}  T. Su, A. Ozcan, On-Chip Holographic Microscopy and its Application for Automated Semen Analysis, Biomed. Opt. Phase Micro. and Nanoscopy, 153-171 (2013).

\bibitem{Mudanyali10} O. Mudanyali, C. Oztoprak, D. Tseng, A. Erlinger, A. Ozcan, Detection of waterborne parasites using field-portable and cost-effective lensfree microscopy, Lab Chip \textbf{10}, 2419-2423 (2010).

\bibitem{Sheng06} J. Sheng, E. Malkiel, J. Katz, Digital Holographic Microscope for Measuring Three-dimensional Particle Distributions and Motions, App. Opt. \textbf{45}, 3893-3901 (2006).

\bibitem{Kiss13} M.Z. Kiss, B.J. Nagy, P. Lakatos, Z. Gorocs, S. Tokes, B. Wittner, L. Orzo, Special Multicolor Illumination and Numerical Tilt Correction in Volumetric Digital Holographic Microscopy, Opt. Exp. \textbf{21}, 12469-12483 (2013).

\bibitem{Mico06} V. Mico, Z. Zalevsky, P. Garcia-Martinez, J. Garcia, Synthetic Aperture Superresolution with Multiple Off-axis holograms, J. Opt. Soc. Am. A \textbf{23}, 3162-3170 (2006).

\bibitem{Mico10} V. Mico, Z. Zalevsky, Superresolved Digital In-line Holographic Microscopy for High-resolution Lensless Biological Imaging, J.  Biomed. Opt. \textbf{15} 046027 (2010).

\bibitem{Granero10} L. Granero, V. Mico, Z. Zalevsky, J. Garcia, Synthetic Aperture Superresolved Microscopy in Digital Lensless Fourier Holography by Time and Angular Multiplexing of the Object Information, App. Opt. \textbf{49}, 845-857 (2010).

\bibitem{Leon08} L. Martinez-Leon, B. Javidi, Synthetic Aperture Single-exposure On-axis Digital Holography, Opt. Exp. \textbf{16}, 161-169 (2008).

\bibitem{Luo15} W. Luo1,.A. Greenbaum, Y. Zhang, A. Ozcan, Synthetic aperture-based on-chip microscopy, Light: Sci. \& App. \textbf{4}, e261 (2015).

\bibitem{Isikman12} S.O. Isikman, A. Greenbaum, W. Luo, A.F. Coskun, A. Ozcan, Giga-Pixel Lensfree Holographic Microscopy and Tomography Using Color Image Sensors PLoS ONE \textbf{7}, e45044 (2012).

\bibitem{Bishra11} W. Bishara, U. Sikora, O. Mudanyali, S. Ting-Wei, O. Yaglidere, S. Luckhart, A. Ozcan, Holographic Pixel Super-resolution in Portable Lensless On-chip Microscopy Using a Fiber-optic Array, Lab Chip \textbf{11}, 1276-1279 (2011).

\bibitem{Greenbaum13}  A. Greenbaum, W. Luo, B. Khademhosseinieh, T. Su, A. Coskun, A. Ozcan, Increased space-bandwidth product in pixel super-resolved lensfree on-chip microscopy, Sci. Rep. \textbf{3}, 1717 (2013).

\bibitem{Isikman10} S.O. Isikman, I. Sencan, O. Mudanyali, W. Bishara, C. Oztopraka, A. Ozcan, Color and Monochrome Lensless On-chip Imaging of Caenorhabditis Elegans Over a Wide Field-of-view, Lab Chip \textbf{10}, 1109-1112 (2010).

\bibitem{Green baum13a} A. Greenbaum, A. Feizi, N. Akbari, A. Ozcan, Wide-field Computational Color Imaging Using Pixel Super-resolved On-chip Microscopy, Opt. Exp. \textbf{21} 12469-12483 (2013).

\bibitem{Noom1} D. Noom, K. Eikema, S. Witte, Lensless phase contrast microscopy based on multiwavlength Fresnel diffraction, Opt. Lett. \textbf{39} 193-196 (2014).

\bibitem{Noom2} D. Noom, D. Flaes, E. Labordus, K. Eikema, S. Witte, High-speed multi-wavelength Fresnel diffraction imaging, Opt. Exp. \textbf{22} 30504-30511 (2014).

\bibitem{Wong15} A. Wong, F. Kazemzadeh, C. Jin, X.Y. Wang, Bayesian-based Aberration Correction and Numerical Diffraction for Improved Lensfree On-chip Microscopy of Biological Specimens, Opt. Lett. \textbf{40}, 2233-2236 (2015).

\bibitem{Kazemzadeh1} F. Kazemzadeh, C. Jin, S. Molladavoodi, Y. Mei, M. B. Emelko, M. B. Gorbet, A Wong, Lens-free spectral light-field fusion microscopy for contrast- and resolution-enhanced imaging of biological specimens, Opt. Lett. \textbf{40}, 3862-3865 (2015).

\bibitem{Kazemzadeh2} F. Kazemzadeh, A. Wong, Lens-free Multi-Laser Spectral Light-Field Fusion Microscopy, Vision Lett. \text{1}, VL102 (2015).

\bibitem{Pelagotti12} A. Pelagotti, M. Paturzo, M. Locatelli, A.  Geltrude, R. Meucci, A. Finizio, P. Ferraro, An Automatic Method for Assembling a Large Synthetic Aperture Digital Hologram, Opt. Exp. \textbf{20}, 4830–4839 (2012).

\bibitem{Ralston07} T. S. Ralston, D. I. Marks, P.S. Carney, S.A. Boppart, Interferometric Synthetic Aperture Microscopy, Nat. Phys. \textbf{3}, 129-134 (2007).

\bibitem{Luo16}  W. Luo, Y. Zhang, A. Feizi, Z. G{\"o}r{\"o}cs, A. Ozcan, Pixel super-resolution using wavelength scanning, Light: Sci. \& App. \textbf{5}, e16060 (2016).

\bibitem{Hell94} S. W. Hell, J. Wichmann, Breaking the diffraction resolution limit by stimulated emission: Stimulated-emission-depletion fluorescence microscopy, Optics Letters \textbf{19} 780-782 (1994).

\bibitem{Betzig92} E. Betzig, J. K. Trautman, Near-Field Optics: Microscopy, Spectroscopy, and Surface Modification Beyond the Diffraction Limit, Science \textbf{257} 189-195 (1992).

\bibitem{Binnig86} G. Binnig, C. F. Quate, C. Gerber, Atomic force microscope, Phys. Rev. Lett. \textbf{56} 930-933 (1986).

\bibitem{Azubel14} M. Azubel, J. Koivisto, S. Malola, D. Bushnell, G. L. Hura, A. L. Koh, H. Tsunoyama, T. Tsukuda, M. Pettersson, H. Häkkinen, R. D. Kornberg, Electron microscopy of gold nanoparticles at atomic resolution, Science \textbf{345} 909-912 (2014).

\bibitem{Mudanyali13} O. Mudanyali, E. McLeod, W. Luo, A. Greenbaum, A. F. Coskun, Y. Hennequin, C. F. Allier, A. Ozcan, Wide-field optical detection of nanoparticles using on-chip microscopy and self-assembled nanolenses, Nat. Phot. \textbf{7} 247-254 (2013).

\bibitem{Polysciences} Polysciences Inc. Fluoresbrite YG Carboxylate Microspheres 0.50$\mu$m. Retrieved from: http://www.polysciences.com/default/fluoresbrite-yg-carboxylate-microspheres-050m (2016).

\end{thebibliography}
\end{document}